\documentclass[aps,showpacs,pra,superscriptaddress,]
{revtex4-2}

\usepackage{amsmath}
\usepackage{amsfonts}
\usepackage{amssymb}
\usepackage{bm}
\usepackage{graphicx}

\begin{document}

\title{Nonlinear Schr\"{o}dinger equations for dense Bose fluid and thin He$^{4}$ film at low temperatures}

\affiliation{Centre for Engineering Quantum Systems, School of Mathematics and Physics, The University of Queensland, Brisbane, Queensland 4072, Australia}
\author{ Vladimir I. Kruglov}

\begin{abstract}
We have derived the novel nonlinear Schr\"{o}dinger equation generalizing the Gross-Pitaevskii (GP) equation for dilute Bose gas. The derivation is based on the Hartree-Fock time-dependent mean-field theory with an arbitrary intermolecular interaction potential. It is shown that obtained nonlinear Schr\"{o}dinger type equation with appropriate redefinition of coefficients can be used for description of the dense Bose fluid at low temperatures. We have obtained the novel nonlinear Schr\"{o}dinger type equation describing the superfluid component of helium in two fluid hydrodynamics. This approach leads to quantum correction for superfluid component of velocity in two fluid hydrodynamics. We have also derived the novel nonlinear Schr\"{o}dinger equation for thin superfluid He$^{4}$ films at low temperatures. It is shown that the Schr\"{o}dinger type equations for dense Bose fluid and thin superfluid He$^{4}$ films lead to phonon-roton dispersion equations for elementary excitation at low temperatures.
\end{abstract}

\maketitle

\section{Introduction}

The study of dynamical properties of
liquid helium at low temperatures (below the lambda
transition) has a long history. A number of methods have been suggested for applications to quantum fluids \cite{A1,A2,A3,A4,A5,A6,A7,A8,A9,A10}. The theory of condensation for a nearly ideal Bose-Einstein
gas is presented, for example, in \cite{A11,A12}. 
It is also shown that at low temperatures and typical pressures the stable clusters in liquid helium consist of 13 bound helium
atoms \cite{A13,A14}, presumably in the form of a central atom surrounded by an icosahedral shell of 12 atoms.
The theory developed in \cite{A15} also yields the ground state energy for the quantum liquid He$^{4}$ in agreement to high accuracy with Monte Carlo simulations and experimental data at low pressure. A new and essential point in this paper is the introduction of a momentum space cutoff at the speed of second sound. In the paper \cite{A16}, the Schr\"{o}dinger equation is solved for approximation of the ground state energies and associated wave functions of carriers confined in a rectangular semiconductor (SC) quantum wire. In this paper
the problem was treated with the effective one band Hamiltonian. 
We note that a model with singular intermolecular interaction should be treated by special methods \cite{B1,B2,B3,B4,B5} that use the Bogoliubov approximation and renormalization procedures to remove the divergences connected with the singular potential. It has been emphasized \cite{B6} that the Hamiltonian with a delta-function potential is pathological and should not be used for correct calculations. The delta-function pseudo-potential was employed by Fermi, but it has been known for a long time that it is not consistent with scattering theory \cite{B7}.
We emphasize that the superfluid He$^{4}$ films offer the significant applications in optomechanics \cite{R1}, neutron storage \cite{R2} and dark matter detection \cite{R3}. The studies of superfluid He$^{4}$ film have also explored different confined geometries  \cite{R4,R5,R6}. The films of He$^{4}$ have attracted a great deal of attention from both theoretical and experimental points of view as they demonstrate superfluidity at extremely low temperatures, which enables for the observation of macroscopic quantum phenomena.
 
The present work consists of the novel Schr\"{o}dinger type equations for dilute Bose gas, dense Bose fluid, superfluid helium in two fluid hydrodynamics and thin He$^{4}$ film at low temperatures. In Sec. II we have derived 
the novel nonlinear Schr\"{o}dinger equation generalizing the Gross-Pitaevskii (GP) equation for dilute Bose gas. The derivation is based on the Hartree-Fock time-dependent mean-field theory with an arbitrary intermolecular interaction potential. We also presented in the Appendix A the modified Born approximation for dilute gas. In Sec. III we presented the generalization of the approach for extended Gross-Pitaevskii equation. In this section, we have derived the novel nonlinear Schr\"{o}dinger type equation describing the liquid He$^{4}$ for dense Bose fluid at low temperatures (below the lambda transition). This approach is based on dispersion relation which leads to the coupling parameters applicable for dense Bose fluid. It is shown that the coefficients in this nonlinear Schr\"{o}dinger equation for dense Bose fluid can be found as the functions of phonon velocity $c$, the wave number $k_{0}$ at roton minimum and the roton minimum energy $\Delta$ of helium Bose fluid. This method yields the unique form for the coefficients in this
nonlinear Schr\"{o}dinger equation. In Sec. IV we have presented the derivation of equations for superfluid component of helium liquid which lead to quantum correction in two fluid hydrodynamics developed by  Landau \cite{Lan,Pat}. This derivation is based on novel nonlinear Schr\"{o}dinger type equation obtained in the Appendix B. This nonlinear Schr\"{o}dinger type equation is derived by the Heisenberg equation for boson annihilation  operator and many-body Hamiltonian for the bosonic system modeling the helium fluid. In the Appendix C we discuss
the Madelung's equations which were obtained for the alternative formulation of the linear Schr\"{o}dinger equation in terms of hydrodynamical variables.
It is shown that the Madelung's equations significantly differ from equations describing the dynamics of superfluid component of density and velocity in the Landau two fluid hydrodynamics with the additional quantum term. In Sec. V we have derived the novel nonlinear Schr\"{o}dinger type equation for thin superfluid He$^{4}$ films. The wave function in this  Schr\"{o}dinger type equation depends on two coordinates describing the position on helium surface. In Sec. VI we present the derivation of dispersion equation for thin superfluid He$^{4}$ films at low temperatures. We
use this dispersion equation in Sec. VII for derivation of the phonon-roton elementary excitations in thin superfluid He$^{4}$ films. We show that the coefficients in this nonlinear Schr\"{o}dinger type equation for thin superfluid He$^{4}$ films are uniquely defined by the third sound velocity $c_{s}$ in helium film, the wave number $k_{0}$ at roton minimum $\Delta$ and the roton minimum energy $\Delta$. The conclusion of this paper is summarized in Section VIII.

\section{Extended GP equation for dilute gas}

In this section, we have derived the novel nonlinear Schr\"{o}dinger type equation generalizing the GP equation for dilute Bose gas. The derivation is based on the Hartree-Fock time-dependent mean-field theory with an arbitrary intermolecular interaction potential.
The modified Hamiltonian $\tilde{H}= H-\mu N$ for many-body Bose system with chemical potential $\mu$ is of the form: 
\begin{equation}
\tilde{H}=-\sum_{j=1}^{N}\frac{\hbar^{2}}{2m}\nabla_{j}^{2}-\mu N
+ \sum_{j<k}^{N}U(|\mathbf{x_{j}}-\mathbf{x_{k}}|).	
\label{1}
\end{equation}
The Hartree-Fock wavefunction describing Bose particles
is given by the product of one-particle wavefunctions as
\begin{equation}
\tilde{\psi}(\mathbf{x_{1}},\mathbf{x_{2}},...,\mathbf{x_{N}};t)=
\prod_{j=1}^{N}	\psi_{N}(\mathbf{x_{j}},t).
\label{2}
\end{equation}
The standard variational procedure with this trial wave
function used in the Hamiltonian (\ref{1}) yields the time
dependent Hartree-Fock equation for the one-particle
wave function $\psi_{N}(x,t)$ as
\begin{equation}
i\hbar\frac{\partial}{\partial t}\psi_{N}(\mathbf{x},t)
=\left\{-\frac{\hbar^{2}}{2m}
\nabla^{2}-\mu+U_{HF}(\mathbf{x},t)\right\}\psi_{N}(\mathbf{x},t), \label{3}
\end{equation}
where $U_{HF}(\mathbf{x},t)$ is the Hartree-Fock time-dependent
mean-field potential given by
\begin{equation}
U_{HF}(\mathbf{x},t)=(N-1)\int_{V} U(|\mathbf{x}-\mathbf{x^{\prime}}|)
|\psi_{N}(\mathbf{x^{\prime}},t)|^{2}d^{3}x^{\prime},	
\label{4}
\end{equation}
with the normalization $\int |\psi_{N}(\mathbf{x},t)|^{2}d^{3}x=1$.

In the case of gas or liquid He$^{4}$ the intermolecular
interactions are given by the Lennard-Jones potential:
\begin{equation}
U(r)=4\epsilon\left[\left(\frac{r_{0}}{r}\right)^{12}
-\left(\frac{r_{0}}{r}\right)^{6}\right].
\label{5}
\end{equation}
The minimum of the potential occurs at $r=r_m=2^{1/6}r_{0}$.
To good accuracy, the parameters of the Lennard-Jones
potential for He$^{4}$ are $\epsilon/k_{B}=10.6\,K$ and $r_{m}=2.98\, \mathring{A}$.
We define an effective interatomic potential $U_{a}(r)$ with the cutoff as: $U_{a}(r)=0$ for $r<a$ and $U_{a}(r)=U(r)$ otherwise. We also introduce the condensate wave function $\Psi(\mathbf{x},t)=\sqrt{N}\psi_{N}(\mathbf{x},t)$. In the thermodynamic limit, when the volume and 
number of particles tend to infinity with fixed local density $n(\mathbf{x},t)\equiv|\Psi(\mathbf{x},t)|^{2}$, the Hartree-Fock equation (\ref{3}) with the changed effective interatomic potential $U_{a}(r)$ 
has the form,
\begin{equation}
i\hbar\frac{\partial}{\partial t}\Psi(\mathbf{x},t)
=\left\{-\frac{\hbar^{2}}{2m}
\nabla^{2}-\mu+\int_{V} U_{a}(|\mathbf{x}-\mathbf{x^{\prime}}|)
|\Psi(\mathbf{x^{\prime}},t)|^{2}
d^{3}x^{\prime}\right\}\Psi(\mathbf{x},t), 
\label{6}
\end{equation}
with the normalization condition for the wavefunction as
$\int_{V} |\Psi(\mathbf{x},t)|^{2}d^{3}x=N$.
Hence, the local density of condensate is $n(\mathbf{x},t)=|\Psi(\mathbf{x},t)|^{2}$.
We can present the density $|\Psi(\mathbf{x^{\prime}},t)|^{2}
=n(\mathbf{x^{\prime}},t)$ in Eq. (\ref{6}) using the Taylor series with standard tensor summation for equal indexes:
\begin{eqnarray}
|\Psi(\mathbf{x^{\prime}},t)|^{2}
=n(\mathbf{x},t)+(\partial_{i}n(\mathbf{x},t))y_{i}
+\frac{1}{2!}(\partial_{i}\partial_{k}n(\mathbf{x},t))y_{i}y_{k}~~~~~~~~~~~~
\nonumber\\ \noalign{\vskip3pt}
+\frac{1}{3!}(\partial_{i}\partial_{k}\partial_{l}n(\mathbf{x},t))
y_{i}y_{k}y_{l}
+\frac{1}{4!}(\partial_{i}\partial_{k}\partial_{l}\partial_{m}n(\mathbf{x},t))y_{i}y_{k}y_{l}y_{m}+R_{5}
(\mathbf{x^{\prime}},t),
\label{7}
\end{eqnarray}
where $n(\mathbf{x},t)=|\Psi(\mathbf{x},t)|^{2}$,  $y_{i}=x_{i}^{\prime}-x_{i}$ and
$\partial_{i}=\partial/\partial x_{i}$  with $i=1,2,3$. 
The remainder $R_{5}(\mathbf{x^{\prime}},t)$ in this Taylor series often is referred to the Lagrange form (or the derivative form). We consider below the case when the remainder  $R_{5}(\mathbf{x^{\prime}},t)$ can be neglected in Eq. (\ref{7}). A reasonable criterion for such approximation
is given in Eq. (\ref{a11}) of Appendix A.  
In this case the integral in Eq. (\ref{6}) is given as
\begin{eqnarray}
\int_{V} U_{a}(|\mathbf{x}-\mathbf{x^{\prime}}|)
|\Psi(\mathbf{x^{\prime}},t)|^{2}d^{3}x^{\prime}
=G_{0}n(\mathbf{x},t)
+G_{i}\partial_{i}n(\mathbf{x},t)~~~~~~~~~~~
\nonumber\\ \noalign{\vskip3pt}
+\frac{1}{2!}G_{ik}\partial_{i}\partial_{k}
n(\mathbf{x},t)+\frac{1}{3!}G_{ikl}\partial_{i}
\partial_{k}\partial_{l}n(\mathbf{x},t)+\frac{1}{4!}
G_{iklm}\partial_{i}\partial_{k}\partial_{l}
\partial_{m}n(\mathbf{x},t),	
\label{8}
\end{eqnarray}
where we use the following notations:
\begin{eqnarray}
G_{0}=\int U_{a}(r)d^{3}y,~~~
G_{i}=\int U_{a}(r)y_{i}d^{3}y,~~~
G_{ik}=\int U_{a}(r)y_{i}y_{k}d^{3}y,
\nonumber\\ \noalign{\vskip3pt}
G_{ikl}=\int U_{a}(r)y_{i}y_{k}y_{l}d^{3}y,~~~
G_{iklm}=\int U_{a}(r)y_{i}y_{k}y_{l}y_{m}
d^{3}y,	  
\label{9}
\end{eqnarray}
with $r=\sqrt{y_{1}^{2}+y_{2}^{2}+y_{3}^{2}}$. The integrals in Eq. (\ref{9}) are defined in infinite volume [with $y_{1}\in (-\infty,+\infty),~ y_{2}\in (-\infty,+\infty),~ y_{3}\in (-\infty,+\infty)$] which is correct for dilute gas because in this case we can consider only two-particle interaction. 
The coupling parameter $G_{0}$ in Eq. (\ref{9}) is given as
\begin{equation}
G_{0}=\int_{-\infty}^{+\infty}\int_{-\infty}^{+\infty}\int_{-\infty}^{+\infty}U_{a}(r)dy_{1}dy_{2}dy_{3}
=4\pi\int_{0}^{\infty}U_{a}(r)
r^{2}dr.
\label{10}
\end{equation}
This equation for the Lennard-Jones potential leads to coupling parameter $G_{0}$ as
\begin{equation}
G_{0}=4\pi\int_{a}^{\infty}U(r)r^{2}dr=\frac{16\pi}{9}\epsilon r_{0}^{3}\left[\left(\frac{r_{0}}{a}\right)^{9}
-3\left(\frac{r_{0}}{a}\right)^{3}\right].
\label{11}
\end{equation}
Hence, the coupling parameter $G_{0}$ tends to infinity when the parameter $a$ tends to zero. Thus, to avoid the divergence we use the cutoff given by finite parameter $a$. We have defined the parameter $a=a_{0}$, where $a_{0}$ is the s-wave scattering length, in the Appendix A.
Equation (\ref{9}) leads to the following relations: $G_{i}=0$ and $G_{ikl}=0$ because the integral obtained for odd function in infinite interval ($-\infty, +\infty$) is zero. The term $G_{ik}\partial_{i}\partial_{k}n$ can written in the form,
\begin{equation}
G_{ik}\partial_{i}\partial_{k}n(\mathbf{x},t)
=G_{11}\partial_{1}^{2}n(\mathbf{x},t)
+G_{22}\partial_{2}^{2}n(\mathbf{x},t)
+G_{33}\partial_{3}^{2}n(\mathbf{x},t)
=G_{11}(\partial_{1}^{2}+\partial_{2}^{2}+
\partial_{3}^{2})n(\mathbf{x},t),
\label{12}
\end{equation}
because $G_{11}=G_{22}=G_{33}$ and
$G_{ik}=0$ for $i\neq k$. The term $G_{iklm}\partial_{i}\partial_{k}\partial_{l}
\partial_{m}n$ is given as
\begin{eqnarray}
G_{iklm}\partial_{i}\partial_{k}\partial_{l}
\partial_{m}n(\mathbf{x},t)=G_{1111}(\partial_{1}^{4}+\partial_{2}^{4}+\partial_{3}^{4})n(\mathbf{x},t)
+6\,G_{1122}(\partial_{1}^{2}\partial_{2}^{2}
+\partial_{1}^{2}\partial_{3}^{2}+\partial_{2}^{2}
\partial_{3}^{2})n(\mathbf{x},t)
\nonumber\\ \noalign{\vskip3pt}
=G_{1111}(\partial_{1}^{2}+\partial_{2}^{2}+
\partial_{3}^{2})^{2}n(\mathbf{x},t)
+q\,(\partial_{1}^{2}\partial_{2}^{2}
+\partial_{1}^{2}\partial_{3}^{2}+\partial_{2}^{2}
\partial_{3}^{2})n(\mathbf{x},t),~~~~~~~~~~~~~~~
\label{13}
\end{eqnarray}
where $q=6G_{1122}-2G_{1111}$. 
We have used in this equation the relations
$G_{1111}=G_{2222}=G_{3333}$ and $G_{1122}=G_{1133}=G_{2233}$.
We neglect below the last term in Eq. (\ref{13}) because the left side in this equation is scalar function and the term $q\,(\partial_{1}^{2}\partial_{2}^{2}
+\partial_{1}^{2}\partial_{3}^{2}+\partial_{2}^{2}
\partial_{3}^{2})n(\mathbf{x},t)$ is not a scalar which is connected with integration in Eq. (\ref{8}) of the Taylor polynomial given by Eq. (\ref{7}) where the remainder $R_{5}
(\mathbf{x^{\prime}},t)=0$. Equation (\ref{8}) with above relations and without the last term in (\ref{13}) has the following form: 
\begin{equation}
\int_{V} U_{a}(|\mathbf{x}-\mathbf{x^{\prime}}|)
|\Psi(\mathbf{x^{\prime}},t)|^{2}d^{3}x^{\prime}
=G_{0}n(\mathbf{x},t)
+\frac{1}{2!}G_{11}\nabla^{2}n(\mathbf{x},t)
+\frac{1}{4!}G_{1111}\nabla^{4}n(\mathbf{x},t),
\label{14}
\end{equation}
where $\nabla^{2}=\partial_{1}^{2}+\partial_{2}^{2}+
\partial_{3}^{2}$ and the operator $\nabla^{4}$ is defined as
\begin{equation}
\nabla^{4}=(\partial_{1}^{2}+\partial_{2}^{2}+
\partial_{3}^{2})^{2}=\partial_{1}^{4}+\partial_{2}^{4}+
\partial_{3}^{4}+2(\partial_{1}^{2}\partial_{2}^{2}
+\partial_{1}^{2}\partial_{3}^{2}+\partial_{2}^{2}
\partial_{3}^{2}).
\label{15}
\end{equation}
The coupling parameters in Eq. (\ref{14}) are $G_{0}=\int U_{a}(r)d^{3}y$, $G_{11}=\int U_{a}(r)y_{1}^{2}d^{3}y$ and $G_{1111}=\int U_{a}(r)y_{1}^{4}d^{3}y$ where $r=\sqrt{y_{1}^{2}+y_{2}^{2}+y_{3}^{2}}$.
Equations (\ref{6}) and (\ref{14}) lead to nonlinear Schr\"{o}dinger equation (NLSE) or extended GP equation for dilute Bose gas:
\begin{equation}
i\hbar\frac{\partial}{\partial t}\Psi(\mathbf{x},t)
=\left\{-\frac{\hbar^{2}}{2m}\nabla^{2}-\mu
+V(|\Psi(\mathbf{x},t)|^{2})\right\}\Psi(\mathbf{x},t), 
\label{16}
\end{equation}
where $\mu=G_{0}|\Psi_{0}|^{2}$ ($|\Psi_{0}|^{2}=n_{0}$) is the chemical potential, and the normalization condition for wave function yields the relation $|\Psi(\mathbf{x},t)|^{2}=n(\mathbf{x},t)$. The nonlinear interaction potential $V(|\Psi(\mathbf{x},t)|^{2})$ in Eq. (\ref{16}) is 
\begin{equation}
V(|\Psi(\mathbf{x},t)|^{2})=G_{0}
|\Psi(\mathbf{x},t)|^{2}
+D_{1}\nabla^{2}|\Psi(\mathbf{x},t)|^{2}
+D_{2}\nabla^{4}|\Psi(\mathbf{x},t)|^{2} ,
\label{17}
\end{equation}
where the parameters $G_{0}$, $D_{1}$ and $D_{2}$ are defined by
the intermolecular interaction potential $U(r)$ as
\begin{equation}
G_{0}=4\pi\int_{a}^{\infty}U(r)r^{2}dr,~~~~
D_{1}=\frac{1}{2}\int U_{a}(r)y_{1}^{2}d^{3}y
=\frac{2\pi}{3}\int_{a}^{\infty} U(r)r^{4}dr,~~~~
D_{2}=\frac{1}{4!}\int U_{a}(r)y_{1}^{4}d^{3}y.
\label{18}
\end{equation}

We note that for many interaction potentials $U(r)$ the integral defining coupling constant $D_{2}$ diverges. Hence, in this case we need more general potential which can be defined as $\tilde{U}(r)=U(r)e^{-\kappa r}$ with additional parameter $\kappa$. However, for dilute gas  the last term $D_{2}\nabla^{4}|\Psi(\mathbf{x},t)|^{2}$ in Eq. (\ref{17}) can be often neglected.
We show in the following Sec. III that the coupling parameters in Eq. (\ref{17}) for dense Bose fluid can be redefined using appropriate dispersion 
equation which leads to the most general form for the nonlinear Schr\"{o}dinger equation (\ref{16}).
Integration in Eq. (\ref{18}) for the Lennard-Jones potential (see Eq. (\ref{5})) leads to the following equations for coupling parameters $G_{0}$ and $D_{1}$: 
\begin{equation}
G_{0}=\frac{16\pi}{9}\epsilon r_{0}^{3}\left[\left(\frac{r_{0}}{a}\right)^{9}
-3\left(\frac{r_{0}}{a}\right)^{3}\right],~~~~
D_{1}=\frac{8\pi}{21}\epsilon r_{0}^{5}
\left[\left(\frac{r_{0}}{a}\right)^{7}
-7\left(\frac{r_{0}}{a}\right)\right],
\label{19}
\end{equation}
where $a$ is the cutoff parameter. We show in Appendix A that $a=a_{0}$ where $a_{0}$ is the s-wave scattering length. Thus, the parameter $a=a_{0}$ is defined (see Eq. (\ref{a9})) as
\begin{equation}
a=\frac{4m\epsilon r_{0}^{3}}{9\hbar^{2}}\left[\left(\frac{r_{0}}{a}\right)^{9}-3\left(\frac{r_{0}}{a}\right)^{3}\right].
\label{20}
\end{equation}

In the conclusion, we note that the derivation of nonlinear Schr\"{o}dinger equation (\ref{16}) with the interaction potential $V(|\Psi(\mathbf{x},t)|^{2})$ given in Eq. (\ref{17}) is based on assumption that the remainder $R_{5}(\mathbf{x^{\prime}},t)$ in Eq. (\ref{7}) can be neglected. The reasonable criterion for the validity of this assumption is presented by Eq. (\ref{a11}) in the Appendix A.
We have presented the energy of elementary excitations  $E_{\mathbf{k}}=\hbar\omega_{\mathbf{k}}$ by Eq. (\ref{a12})  which follows from extended GP equation (\ref{16}). Equation (\ref{a12}) demonstrates that the parameters $G_{0}$, $D_{1}$ and $D_{2}$  define the terms proportional to $k^{2}$, $k^{4}$ and $k^{6}$ for the function $\omega_{\mathbf{k}}^{2}$ respectively. The parameters $G_{0}$, $D_{1}$ and $D_{2}$ also determine the nonlinear interaction potential $V(|\Psi(\mathbf{x},t)|^{2})$ in Eq. (\ref{17}). Moreover, Eq. (\ref{18}) shows that these parameters are completely defined by the intermolecular interaction potential $U(r)$ and the cutoff parameter $a$ or 
s-wave scattering length $a_{0}$.

\section{Nonlinear Schr\"{o}dinger equation for dense Bose fluid}

In this section, we have derived the novel nonlinear Schr\"{o}dinger type equation describing the liquid He$^{4}$ for dense Bose fluid at low temperatures (below the lambda transition). This approach is based on derivation of generalized dispersion relation which leads to the coupling parameters applicable for dense Bose fluid. It is shown that the coefficients in this nonlinear Schr\"{o}dinger equation for dense Bose fluid can be found as the fuctions depending on phonon velocity $c$, the wave number $k_{0}$ at roton minimum and the roton minimum energy $\Delta$ of helium Bose fluid. This method yields the unique form for the coefficients in this nonlinear Schr\"{o}dinger equation.
We emphasize that the approach and approximation used in this paper are valid in the dense-fluid regime as well.    

We note that the nonlinear Schr\"{o}dinger type equation describing a dense Bose fluid at low temperatures 
has the same form as Eqs. (\ref{16}) and (\ref{17}). However,
in this case the coefficients in Eq. (\ref{17}) are connected
with many-body interaction. Thus, the generalized nonlinear Schr\"{o}dinger equation describing a dense Bose fluid follows
from Eqs. (\ref{16}) and (\ref{17}) with the change $G_{0}\mapsto F_{0}$, $D_{1}\mapsto F_{1}$ and $D_{2}\mapsto F_{2}$ in Eq. (\ref{17}):
\begin{equation}
i\hbar\frac{\partial}{\partial t}\Psi(\mathbf{x},t)
=\left\{-\frac{\hbar^{2}}{2m}\nabla^{2}-\mu
+F_{0}|\Psi|^{2}
+F_{1}\nabla^{2}|\Psi|^{2}
+F_{2}\nabla^{4}|\Psi|^{2}\right\}\Psi(\mathbf{x},t). 
\label{21}
\end{equation}
We can define the coupling parameters $F_{n}$ as the functions which depend on the phonon velocity $c$, the wave number $k_{0}$ at roton minimum and the roton minimum energy $\Delta$
of helium Bose fluid. These parameters $c$, $k_{0}$ and $\Delta$ are uniquely connected with the energy spectrum of the elementary excitation in a dense Bose fluid. 
We describe the elementary excitations in dense He$^{4}$ Bose fluid using the stationary solution $\Psi_{0}=\sqrt{n_{0}}$, chemical potential $\mu=F_{0}n_{0}$ and small perturbation $\Psi^{\prime}(\mathbf{x},t)$ as  
\begin{equation}
\tilde{\Psi}(\mathbf{x},t)=\sqrt{n_{0}}
+\Psi^{\prime}(\mathbf{x},t),~~~~
\Psi^{\prime}(\mathbf{x},t)=
A_{\mathbf{k}}e^{i(\mathbf{k}\mathbf{x}
-\omega_{\mathbf{k}}t)}+B^{\ast}_{\mathbf{k}}
e^{-i(\mathbf{k}\mathbf{x}-\omega_{\mathbf{k}}t)},	\label{22}
\end{equation}
where $\mathbf{k}$ is the wave number. The linearization of Eq. (\ref{21}) to small fluctuations $\Psi^{\prime}(\mathbf{x},t)$ yields the system of equations,
\begin{equation}
[E_{\mathbf{k}}-\epsilon_{\mathbf{k}}
-n_{0}W(k)]A_{\mathbf{k}}
-n_{0}W(k)B_{\mathbf{k}}=0,
\label{23}
\end{equation}
\begin{equation}
[E_{\mathbf{k}}+\epsilon_{\mathbf{k}}+n_{0}W(k)]
B_{\mathbf{k}}+n_{0}W(k)A_{\mathbf{k}}=0,  
\label{24}
\end{equation}
where $E_{\mathbf{k}}=\hbar\omega_{\mathbf{k}}$ and $\epsilon_{\mathbf{k}}=\frac{\hbar^{2} k^{2}}{2m}$ are the energy of elementary excitation and free particle energy respectively. The function $W(k)$ is given in these equations by
\begin{equation}
W(k)=F_{0}-F_{1}k^{2}+F_{2}k^{4}.
\label{25}
\end{equation}
The non-zero solution of Eqs. (\ref{23}) and (\ref{24}) leads to the energy of elementary excitation in dense He$^{4}$ Bose fluid:
\begin{equation}
E_{\mathbf{k}}=\left[\left(\frac{\hbar^{2}k^{2}}{2m}\right)^{2}+n_{0}\left(\frac{\hbar^{2} k^{2}}{m}\right)
(F_{0}-F_{1}k^{2}+F_{2}k^{4})\right]^{1/2}.
\label{26}
\end{equation}
We can also write Eq. (\ref{26}) in the form, 
\begin{equation}
\varepsilon^{2}(p)=\lambda_{1}p^{2}+\lambda_{2}p^{4} +\lambda_{3}p^{6},
\label{27}
\end{equation}
with $\varepsilon(p)=E(k)$, $p=\hbar k$, and the parameters $\lambda_{n}$ are given as 
\begin{equation}
\lambda_{1}=\frac{n_{0}F_{0}}{m},~~~~\lambda_{2}
=\frac{1}{4m^{2}}-\frac{n_{0}F_{1}}{m\hbar^{2}},~~~~
\lambda_{3}=\frac{n_{0}F_{2}}{m\hbar^{4}}.	
\label{28}
\end{equation}
We have the following conditions for the coefficients $\lambda_{n}$ depending on photon velocity $c$, momentum $p_{0}=\hbar k_{0}$ and
 the roton minimum energy $\Delta$:
\begin{equation}
\left(\frac{d\varepsilon(p)}{dp}\right)_{p=0}=c, ~~~~ \left(\frac{d\varepsilon(p)}{dp}\right)_{p=p_{0}}=0,~~~~ \varepsilon(p_{0})=\Delta.
\label{29}
\end{equation}
These conditions lead to the following equations for parameters $\lambda_{n}$: 
\begin{equation}
\sqrt{\lambda_{1}}=c,~~~~\lambda_{1}+2\lambda_{2}p_{0}^{2}+3\lambda_{3}p_{0}^{4}=0,  
\label{30}
\end{equation}
\begin{equation}
\lambda_{1}p_{0}^{2}+\lambda_{2}p_{0}^{4} +\lambda_{3}p_{0}^{6}=\Delta^{2}.
\label{31}
\end{equation}
Thus, we have the coefficients $\lambda_{n}$ in the form: 
\begin{equation}
\lambda_{1}=c^{2},~~~~ \lambda_{2}=\frac{3\Delta^{2}}{p_{0}^{4}} 
-\frac{2c^{2}}{p_{0}^{2}},~~~~ \lambda_{3}=\frac{c^{2}}{p_{0}^{4}} 
-\frac{2\Delta^{2}}{p_{0}^{6}},  
\label{32}
\end{equation}
where $\lambda_{3}>0$, $\lambda_{2}<0$ and $\lambda_{1}>0$. Equations (\ref{28}) and (\ref{32}) lead to the following parameters $F_{n}$:
\begin{equation}
F_{0}=\frac{mc^{2}}{n_{0}},~~~~ F_{1}=\frac{\hbar^{2}}{4mn_{0}} 
+\frac{2m{c}^{2}}{n_{0}k_{0}^{2}} -\frac{3m\Delta^{2}}{n_{0}\hbar^{2}k_{0}^{4}},~~~~
F_{2}=\frac{mc^{2}}{n_{0}k_{0}^{4}} -\frac{2m\Delta^{2}}{n_{0}\hbar^{2}k_{0}^{6}}.   
\label{33}
\end{equation}
The phonon velocity $c$ also follows from the standard definition,
\begin{equation}
c=\lim_{k\rightarrow 0}=\frac{1}{\hbar}
\frac{\partial E_{\mathbf{k}}}{\partial k}=\sqrt{\frac{n_{0}\,F_{0}}{m}},
\label{34}
\end{equation}
which yields the relation $F_{0}=mc^{2}/n_{0}$. 
We note that the sound velocity in the
liquid He$^{4}$ at zero temperature is to high accuracy a linear function of the mass density $\rho=nm$. Thus, the sound velocity $c$ in the liquid He$^{4}$ is given by the following relation \cite{A15}:
\begin{equation}
c=\sigma_{0}+\sigma_{1}mn_{0},
\label{35}
\end{equation}
where the parameters $\sigma_{0}=-4.116\times 10^{4}$ and $\sigma_{1}=4.473\times 10^{4}$ (in c.g.s. units) are found from experimental and numerical data.

\section{Quantum correction for two fluid hydrodynamics}

In this section, we present the derivation of equations for superfluid helium liquid with quantum correction in two fluid hydrodynamics developed by  Landau \cite{Lan,Pat}. This derivation is based on the novel nonlinear Schr\"{o}dinger type equation obtained in the Appendix B. This nonlinear Schr\"{o}dinger type equation is derived by the Heisenberg equation for boson annihilation field operator in a Fock space with many-body Hamiltonian for the bosonic system modeling the helium fluid.

We define the boson annihilation operator $\hat{\Psi}(\mathbf{x},t)$ at Heisenberg representation in the following form,
\begin{equation}
\hat{\Psi}(\mathbf{x},t)=\Psi(\mathbf{x},t)+\hat{\eta}
(\mathbf{x},t),~~~~\langle\hat{\Psi}^{\dagger}
(\mathbf{x},t)\hat{\Psi}(\mathbf{x},t)\rangle
=\nu_{s}(\mathbf{x},t)+\nu_{n}(\mathbf{x},t),
\label{36}
\end{equation}
where the wave function $\Psi(\mathbf{x},t)$ and the Heisenberg operator $\hat{\eta}(\mathbf{x},t)$ are
defined by the following conditions: $\langle\hat{\Psi}(\mathbf{x},t)\rangle=\Psi(\mathbf{x},t)$
and $\langle\hat{\eta}(\mathbf{x},t)\rangle=0$. Here $\langle...\rangle=$Tr$(...\rho_{0})$ where $\rho_{0}$ is
the density operator at an initial time $t=0$. 
The superfluid and normal densities of helium fluid below lambda transition are given
as $\nu_{s}(\mathbf{x},t)=|\Psi(\mathbf{x},t)|^{2}$ and $\nu_{n}(\mathbf{x},t)=\langle\hat{\eta}^{\dagger}
(\mathbf{x},t)\hat{\eta}(\mathbf{x},t)\rangle$.
The nonlinear Schr\"{o}dinger type equation for the wave function $\Psi(\mathbf{x},t)$ is given (see the Appendix B) in the following form:
\begin{equation}
i\hbar\frac{\partial}{\partial t}\Psi(\mathbf{x},t)
=\left\{-\frac{\hbar^{2}}{2m}\nabla^{2}-\mu
+\langle\hat{U}(\mathbf{x},t)\rangle\right\}
\Psi(\mathbf{x},t),
\label{37}
\end{equation}
where the potential energy $\langle\hat{U}(\mathbf{x},t)\rangle$ is given as
\begin{equation}
\langle\hat{U}(\mathbf{x},t)\rangle=\int U(|\mathbf{x}-\mathbf{x^{\prime}}|)(\nu_{s}
(\mathbf{x}^{\prime},t)+\nu_{n}(\mathbf{x}^{\prime},t))
d^{3}x^{\prime}.
\label{38}
\end{equation}
Here $\nu_{s}$ and $\nu_{n}$ are the superfluid and normal  density respectively.
The wave function of helium superfluid can be written in the form $\Psi(\mathbf{x},t)=\sqrt{n_{s}(\mathbf{x},t)} 
\exp[i\Phi(\mathbf{x},t)]$. This wave function leads to mass density $\rho_{s}$ and velocity $\mathbf{v}_{s}$ for superfluid helium component: 
\begin{equation}
\rho_{s}=m|\Psi(\mathbf{x},t)|^{2},~~~~
\mathbf{v}_{s}=\frac{\hbar}{m}\nabla\Phi(\mathbf{x},t).
\label{39}
\end{equation}
The nonlinear Schr\"{o}dinger type equation (\ref{37}) and relations given in Eq. (\ref{39}) lead to the following system of equations: 
\begin{equation}
\partial_{t}\rho_{s}+\nabla\cdot (\rho_{s}\mathbf{v}_{s})=0,
\label{40}
\end{equation}
\begin{equation}
\hbar\partial_{t}\Phi+\frac{1}{2}m\mathbf{v}_{s}^{2}
+\langle\hat{U}(\mathbf{x},t)\rangle-\mu=\frac{\hbar^{2}}{2m\sqrt{\rho_{s}}}\nabla^{2}\sqrt{\rho_{s}},
\label{41}
\end{equation}
We can apply the operator $(1/m)\nabla)$ to Eq. (\ref{41}) wich yields the hydrodynamic equation for superfluid velocity,
\begin{equation}
\partial_{t}\mathbf{v}_{s}+\nabla \left(\frac{1}{2}\mathbf{v}_{s}^{2}+\frac{1}{m}\mu_{0}
(\mathbf{x},t)\right)
=\nabla R,~~~~R=\frac{\hbar^{2}}{2m^{2}\sqrt{\rho_{s}}}
\nabla^{2}\sqrt{\rho_{s}},
\label{42}
\end{equation}
where $\mu_{0}(\mathbf{x},t)=\langle\hat{U}(\mathbf{x},t)\rangle-\mu$. We note that Eq. (\ref{42}) has also a quantum correction term $\nabla R$ given in the right side of this
equation. We should consider the influence of normal component of helium fluid on superfluid velocity $\mathbf{v}_{s}$.   
For this reason we take into account the conservation equation for energy used in two fluid hydrodynamics:  $\partial_{t}E+\nabla\cdot \mathbf{Q}=0$ \cite{Lan,Pat}, where $E$ is the energy for unit of volume and $\mathbf{Q}$ is the flow of energy. This conservation equation yields for the redefined chemical potential $\tilde{\mu}$ the following equations:
\begin{equation}
\frac{1}{m}\mu_{0}(\mathbf{x},t)=\tilde{\mu}(\mathbf{x},t),
~~~~
d\tilde{\mu}=\frac{1}{\rho}dp-\frac{1}{\rho}SdT
-\frac{\rho_{n}}{2\rho}d\mathbf{w}^{2},
\label{43}
\end{equation}
where $\mathbf{w}=\mathbf{v}_{n}-\mathbf{v}_{s}$, $p$ and $S$ are the pressure and entropy of helium fluid respectively, and
$\mathbf{v}_{n}$ is the velocity of normal component
of helium fluid. Equation (\ref{43}) leads to the following relation:
\begin{equation}
\nabla\tilde{\mu}=\frac{1}{\rho}\nabla p
-\frac{1}{\rho}S\nabla T-\frac{\rho_{n}}{2\rho}\nabla\mathbf{w}^{2}.
\label{44}
\end{equation}
It follows from Eq. (\ref{39}) that $\mathrm{curl}\,\mathbf{v}_{s}=0$ because $\mathrm{curl}\,\nabla \Phi\equiv 0$. Thus, we have the relation $(1/2)\nabla\mathbf{v}_{s}^{2}
=(\mathbf{v}_{s}\cdot \nabla)\mathbf{v}_{s}$. Equations (\ref{42}) and (\ref{44}) yield the equation for superfluid velocity with quantum correction term as
\begin{equation}
\partial_{t}\mathbf{v}_{s}
+(\mathbf{v}_{s}\cdot\nabla)\mathbf{v}_{s}
=-\frac{1}{\rho}\nabla p
+\frac{1}{\rho}S\nabla T
+\frac{\rho_{n}}{2\rho}\nabla\mathbf{w}^{2}+\nabla R.
\label{45}
\end{equation}

The complete system of two fluid hydrodynamics contains also the conservation equations for the  mass and momentum of fluid and the equation for entropy $S$. The conservation equation for the mass of fluid is
\begin{equation}
\partial_{t}\rho+\nabla\cdot \mathbf{J}=0,
\label{46}
\end{equation}
where $\rho=\rho_{s}+\rho_{n}$, and $\mathbf{J}$ is the momentum of unit volume given by
\begin{equation}
\mathbf{J}=\rho_{s}\mathbf{v}_{s}+\rho_{n}\mathbf{v}_{n}.
\label{47}
\end{equation}
Here $\rho_{n}$ and $\mathbf{v}_{n}$ are the mass density and the velocity for normal component of helium fluid.
The conservation equation for the momentum of fluid
has the standard form:
\begin{equation}
\partial_{t}J_{i}+\partial_{k}\Pi_{ik}=0,
\label{48}
\end{equation}
where $\Pi_{ik}$ is the tensor of momentum flow which is given in two fluid hydrodynamics as 
\begin{equation}
\Pi_{ik}=\rho_{s}v_{si}v_{sk}+\rho_{n}v_{ni}v_{nk}
+p\delta_{ik}.
\label{49}
\end{equation}
The equation for entropy $S$ has the following form:
\begin{equation}
\partial_{t}S+\nabla\cdot\left(S\mathbf{v}_{n}
-\frac{\kappa}{T}\nabla T\right)=\frac{\kappa}{T^{2}}
(\nabla T)^{2},
\label{50}
\end{equation}
where $\kappa$ is the thermal conductivity. This parameter
is given as $\kappa=\kappa_{p}+\kappa_{r}$ where
$\kappa_{p}$ and $\kappa_{r}$ are the phonon and roton thermal conductivity respectively \cite{R7}. 

We discuss in the Appendix C
the Madelung's equations obtained for alternative formulation of the linear Schr\"{o}dinger equation for one particle wave function. It is shown in this Appendix C that the Madelung's equations significantly differ from equations describing the dynamics of superfluid component of density and velocity in the Landau two fluid hydrodynamics with additional quantum term $\nabla R$ presented in the right side of Eq. (\ref{45}). 

\section{Nonlinear Schr\"{o}dinger equation for thin superfluid He$^{4}$ films}
 
In this section, we present the derivation of two-dimensional Schr\"{o}dinger type equation for thin superfluid He$^{4}$ films.
The theory predicts the phonon-roton elementary excitation spectrum for thin helium film. We define the complex wave function $\psi(x,y,t)$ (or complex order parameter) which is a function of two coordinates $x$ and $y$ describing position on the surface.
In thin-film limit the surface tension falls to zero and the surface rotons are
characterized by an energy gap of $5.3$ K. Moreover, for thin films with a superfluid surface density of order one or few atomic layers the GP healing concept is not applicable at any temperature Ref. \cite{Rut}. The quantum description of thin helium films by nonlinear two-dimensional Schr\"{o}dinger equation takes place when the condition $k_{0}\zeta_{0}\lesssim 1$ is satisfied where $k_{0}$ is the wave number at surface roton minimum. Here $\zeta_{0}$ is the non-perturbed thickness of superfluid layer and the remaining thickness $d-\zeta_{0}$ is connected with the normal component of He$^{4}$. We note that the above condition leads to the following inequalities $k\zeta_{0}< 1$ or $k\zeta_{0}\ll 1$ (for enough small $k$) where $k$ is the wave number of the third-sound.
The discussion of conditions for two-dimensional quantum hydrodynamics in very thin helium films is presented in Ref. \cite{Rut}. However, the novel nonlinear Schr\"{o}dinger type equation derived in this section  differs from the proposed in above paper a theoretical model for superfluidity in thin helium films.

Thus, we assume that discussed above conditions are satisfied
for thin superfluid He$^{4}$ films. In this case the wave function $\psi_{N}(x_{1},x_{2},t)$ in Hartree-Fock equation (\ref{3}) depends on two coordinates and the superfluid thickness in helium film is given by the function $\zeta(x_{1},x_{2},t)$ as 
\begin{equation}
\zeta(x_{1},x_{2},t)=\zeta_{0}+\eta(x_{1},x_{2},t),
\label{51}
\end{equation}
where $\zeta_{0}$ is the non-perturbed thickness of the superfluid helium layer of the film. We formulate the normalization condition for the wave function $\psi_{N}(x_{1},x_{2},t)$ in two dimensional space $(x_{1},x_{2})\in \mathcal{S}$ because this wave function does not depend on the third (vertical) coordinate $x_{3}$. The volume $V$ for non-perturbed helium film is $V=S\times \zeta_{0}$, hence the normalization condition $\int_{V}|\psi_{N}(x_{1},x_{2},t)|^{2}d^{3}x=1$ for the non-perturbed helium film is
$\int_{S}|\psi_{N}(x_{1},x_{2},t)|^{2}
\zeta_{0}dx_{1}dx_{2}=1$.
In general case the normalization condition for the wave function $\psi_{N}(x_{1},x_{2},t)$ of helium film is
\begin{equation}
\int_{S}|\psi_{N}(x_{1},x_{2},t)|^{2}dx_{1}dx_{2}
=\frac{1}{\zeta_{0}},
\label{52}
\end{equation}
where the right side in this equation is a constant. The interatomic potential $U_{a}(r)$ with the cutoff is given as: $U_{a}(r)=0$ for $r<a$ and $U_{a}(r)=U(r)$ otherwise. We introduce also the condensate wave function by the following relation $\Psi(x_{1},x_{2},t)=\sqrt{N}\psi_{N}(x_{1},x_{2},t)$.
Equation for the wave function $\Psi(x_{1},x_{2},t)$ follows from Eq. (\ref{21}) with the change of wave function $\Psi(\mathbf{x},t)\mapsto \Psi(x_{1},x_{2},t)$ and coefficients $F_{n}\mapsto \tilde{F_{n}}$.
Thus, we have the following equation for the wave function $\Psi(x_{1},x_{2},t)\equiv\Psi(x,y,t)$ (with $x_{1}=x$ and $x_{2}=y$) which does not depend on the variable $x_{3}$:
\begin{equation}
i\hbar\frac{\partial}{\partial t}\Psi(x,y,t)
=\left\{-\frac{\hbar^{2}}{2m}\nabla_{s}^{2}-\mu
+V(|\Psi|^{2})\right\}
\Psi(x,y,t), 
\label{53}
\end{equation}
where the nonlinear potential $V(|\Psi|^{2})$ is given as
\begin{equation}
V(|\Psi|^{2})=\tilde{F_{0}}|\Psi(x,y,t)|^{2}
+\tilde{F_{1}}\nabla_{s}^{2}|\Psi(x,y,t)|^{2}	+\tilde{F_{2}}\nabla_{s}^{4}|\Psi(x,y,t)|^{2},	
\label{54}
\end{equation}
with $\nabla_{s}^{2}=\partial_{x}^{2}+\partial_{y}^{2}$,
and $\nabla_{s}^{4}=(\partial_{x}^{2}
+\partial_{y}^{2})^{2}=\partial_{x}^{4}+\partial_{y}^{4}
+2\partial_{x}^{2}\partial_{y}^{2}$. The chemical potential
in Eq. (\ref{53}) is $\mu=\tilde{F_{0}}|\Psi_{0}|^{2}$. 
We note that in Eq. (\ref{54}) the coefficients $\tilde{F_{n}}$ for superfluid He$^{4}$ films differ from appropriate coefficients $F_{n}$ obtained for dense Bose fluid in Sec. III. We derive the explicit form for these coefficients $\tilde{F_{n}}$ in Sec. VII
using the dispersion equation for superfluid He$^{4}$ films obtained in Sec. VI.

The normalization condition for the wave function
$\Psi(x,y,t)$ follows from Eq. (\ref{52}) and
relation $\Psi(x,y,t)=\sqrt{N}\psi_{N}(x,y,t)$:
\begin{equation}
\int_{S}|\Psi(x,y,t)|^{2}dxdy
=\frac{N}{\zeta_{0}}.  
\label{55}
\end{equation}
Moreover, we have the equation $\int_{S}n_{0}\zeta(x,y,t)dxdy
=N$ where $n_{0}\zeta(x,y,t)$ is the unique function in the integral leading to number $N$ of atoms in the helium film for given density $n_{0}$ and an arbitrary surface shape of helium film defined by the function $\zeta(x,y,t)$. The comparison of this equation and (\ref{55}) yields the relation:
\begin{equation}
|\Psi(x,y,t)|^{2}=\frac{n_{0}}{\zeta_{0}}
\zeta(x,y,t). 
\label{56}
\end{equation}
We also define the wave function:
$\psi(x,y,t)=\sqrt{\zeta_{0}/n_{0}}
\Psi(x,y,t)$ which yields the following relation 
$|\Psi(x,y,t)|^{2}=(n_{0}/\zeta_{0})|\psi(x,y,t)|^{2}$. This relation with Eqs. (\ref{55}) and (\ref{56}) leads to normalization condition for the wave function $\psi(x,y,t)$ as
\begin{equation}
\int_{S}n_{0}|\psi(x,y,t)|^{2}dxdy
=N,~~~~|\psi(x,y,t)|^{2}
=\zeta(x,y,t).
\label{57}
\end{equation}
The substitution of wave function  $\Psi(x,y,t)=\sqrt{n_{0}/\zeta_{0}}\,
\psi(x,y,t)$ and $\Psi_{0}=\sqrt{n_{0}/\zeta_{0}}\,
\psi_{0}$ to Eqs. (\ref{53}) and (\ref{54}) leads to the
nonlinear Schr\"{o}dinger type equation for superfluid He$^{4}$ films:
\begin{equation}
i\hbar\frac{\partial}{\partial t}\psi(x,y,t)
=\left\{-\frac{\hbar^{2}}{2m}\nabla_{s}^{2}-\mu+V_{a}
(|\psi|^{2})\right\}\psi(x,y,t). 
\label{58}
\end{equation}
The nonlinear potential $V_{a}(|\psi|)$ in this equation is given by
\begin{equation}
V_{a}(|\psi|^{2})=G
|\psi(x,y,t)|^{2}
+\beta\nabla_{s}^{2}|\psi(x,y,t)|^{2}	+\sigma\nabla_{s}^{4}|\psi(x,y,t)|^{2},
\label{59}
\end{equation}
where the coefficients $G$, $\beta$ and $\sigma$ are defined as
\begin{equation}
G=\frac{n_{0}}{\zeta_{0}}\tilde{F_{0}},~~~~
\beta=\frac{n_{0}}{\zeta_{0}}\tilde{F_{1}},~~~~
\sigma=\frac{n_{0}}{\zeta_{0}}\tilde{F_{2}},
\label{60}
\end{equation}
and the chemical potential in Eq. (\ref{58}) is given by standard relation $\mu=G|\psi_{0}|^{2}$.

\section{Dispersion equation for thin superfluid He$^{4}$ films}

In this section we obtain the dispersion equation for thin superfluid He$^{4}$ films at low temperatures.
We use one-dimensional Schr\"{o}dinger type equation for derivation of the dispersion equation in thin superfluid He$^{4}$ films because two-dimensional Schr\"{o}dinger type equation (\ref{58}) leads to the same result as the one-dimensional case. 
Nonlinear Schr\"{o}dinger equation (\ref{58}) for superfluid He$^{4}$ films in one-dimensional case 
has the following form: 
\begin{equation}
i\hbar\frac{\partial\psi}{\partial t}=-\frac{\hbar^{2}}{2m}\frac{\partial
^{2}\psi}{\partial x^{2}}+ G\left(|\psi|^{2}-|\psi_{0}|^{2}\right)\psi+\beta%
\frac{\partial ^{2}|\psi|^{2}}{\partial x^{2}}\psi +\sigma\frac{\partial
^{4}|\psi|^{2}}{\partial x^{4}}\psi.  
\label{61}
\end{equation}
We present the wave function in the form $\psi(x,t)=U(x,t)\exp(i\Theta(x,t)/\hbar)$ with $U(x,t)=\sqrt{\zeta(x,t)}$ then the nonlinear Schr\"{o}dinger Eq. (\ref{61}) yields the system of equations: 
\begin{equation}
U_{t}=-\frac{1}{2m}\Theta_{xx}U-\frac{1}{m}
\Theta_{x}U_{x},  
\label{62}
\end{equation}
\begin{equation}
-\Theta_{t}=-\frac{\hbar^{2}}{2m}\frac{(\zeta^{1/2})_{xx}} {\zeta^{1/2}}+%
\frac{1}{2m} (\Theta_{x})^{2}+G(\zeta-\zeta_{0})+\beta\zeta_{xx}
+\sigma
\zeta_{xxxx}.  
\label{63}
\end{equation}
Equation (\ref{62}) can also be written as 
\begin{equation}
\zeta_{t}+\frac{1}{m}(\Theta_{x}\zeta)_{x}=0, 
\label{64}
\end{equation}
which is the continuity equation.
We use the standard definition of helium velocity in a film as $v_{k}=(1/m)\partial_{k}\Theta$ where $k=1,2$ and $v_{1}=u,\, v_{2}=w$, then the velocity is $u=m^{-1}\Theta_{x}$ because for one-dimensional case $w=0$. Equation (\ref{64}) leads to the following form of continuity equation: 
\begin{equation}
\zeta_{t}+(u\zeta)_{x}=0.  
\label{65}
\end{equation}

We can write the thickness for superfluid component of helium and the phase as $\zeta=\zeta_{0}+\eta$ and $\Theta=\Theta_{0}+\theta$ where $\eta$ is the deviation of surface for superfluid helium component from non exited depth $\zeta_{0}=\mathrm{const}$. The parameter  $\Theta_{0}=\mathrm{const}$ is a phase for ground state of helium in the film and $\theta$ is a deviation of the phase. Linearization of the system of Eqs. (\ref{63}) and (\ref{64}) to small deviations for $\eta$ and $\theta$ leads to the following system of linear equations: 
\begin{equation}
-\theta_{t}=-\frac{\hbar^{2}}{4m\zeta_{0}}\eta_{xx}
+G\eta+\beta\eta_{xx}+\sigma\eta_{xxxx},  
\label{66}
\end{equation}
\begin{equation}
\eta_{t}=-\frac{\zeta_{0}}{m}\theta_{xx}.  
\label{67}
\end{equation}
The system of Eqs. (\ref{66}) and (\ref{67}) has the following solution: 
\begin{equation}
\eta=a(k)\cos(kx-\omega t),~~~~ 
\theta=b(k)\sin(kx-\omega t),  
\label{68}
\end{equation}
with the wave number $k$ and frequency $\omega$. The substitution of functions given in Eq. (\ref{68}) to Eqs. (\ref{66}) and (\ref{67}) yields the system of equations as 
\begin{equation}
b(k)\omega=\frac{\hbar^{2}a(k)k^{2}}{4m\zeta_{0}}+Ga(k) -\beta a(k)k^{2}+\sigma a(k)k^{4},~~~~
a(k)\omega=\frac{\zeta_{0}}{m}b(k)k^{2}.
\label{69}
\end{equation}
The solution of these linear equations leads to dispersion equation for superfluid He$^{4}$ films at low temperatures: 
\begin{equation}
\omega^{2}=\frac{G}{m\zeta_{0}}(k\zeta_{0})^{2} +\left(\frac{\hbar^{2}}{%
4m^{2}\zeta_{0}^{4}}-\frac{\beta} {m\zeta_{0}^{3}}\right)(k\zeta_{0})^{4} +%
\frac{\sigma}{m\zeta_{0}^{5}}(k\zeta_{0})^{6}.  
\label{70}
\end{equation}
We use this dispersion relation in the next section for derivation of equation for the phonon-roton elementary excitations in thin superfluid He$^{4}$ films. The
obtained dispersion relation shows that the coefficients $G$, $\beta$ and $\sigma$ are connected with the terms proportional to 
$k^{2}$, $k^{4}$ and $k^{6}$ for the function $\omega^{2}$ respectively.

\section{Phonon-roton elementary excitations in superfluid He$^{4}$ films}

In this section, we define the coefficients $G$, $\beta$ and $\sigma$ in nonlinear Schr\"{o}dinger equation (\ref{58}) using the third sound velocity $c_{s}$ in helium film,
the wave number $k_{0}$ at roton minimum and the roton minimum energy $\Delta$.  It is shown that the coefficients in this novel Schr\"{o}dinger type nonlinear equation for thin superfluid He$^{4}$ films are uniquely defined by the third sound velocity $c_{s}$ in helium film, the wave number $k_{0}$ at roton minimum and the roton minimum energy $\Delta$.
The parameters $k_{0}$ and $\Delta$ are connected by relation $E(k_{0})=\Delta$ where $E(k)$ is the energy of elementary excitations in superfluid He$^{4}$ films. Dispersion equation (\ref{70}) for superfluid He$^{4}$ films at low temperatures can also be written: 
\begin{equation}
E(k)=(b_{1}k^{2}+b_{2}k^{4} +b_{3}k^{6})^{1/2},  
\label{71}
\end{equation}
where $E(k)=\hbar\omega(k)$ and the coefficients $b_{n}$ are given as
\begin{equation}
b_{1}=\frac{G\zeta_{0}\hbar^{2}}{m},~~~~b_{2} =\frac{\hbar^{4}}{4m^{2}}-\frac{\beta\zeta_{0}\hbar^{2}}{m},~~~~b_{3}=\frac{\sigma\zeta_{0}\hbar^{2}}{m}.  
\label{72}
\end{equation}
\begin{figure}[h]
\includegraphics[width=1\textwidth]{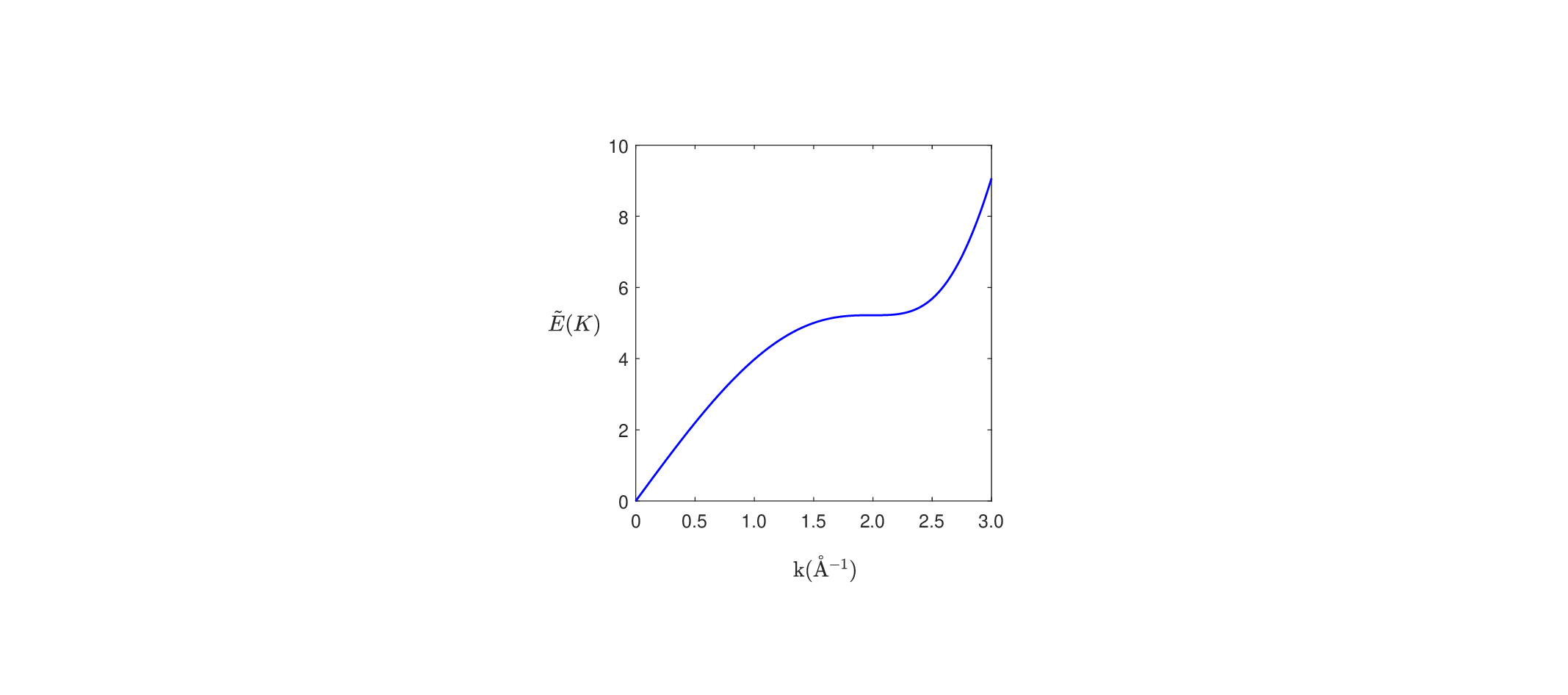}
\caption{Phonon-roton dispersion relation for elementary excitation in superfluid He$^{4}$ film given by Eq. (\ref{81}) for parameters: $c_{s}=59.36~m/s$, $\Delta /k_{B}=5.22$ K and $k_{0}=2~\mathring{A}^{-1}$. }
\label{FIG.1.}
\end{figure}
\begin{figure}[h]
\includegraphics[width=1\textwidth]{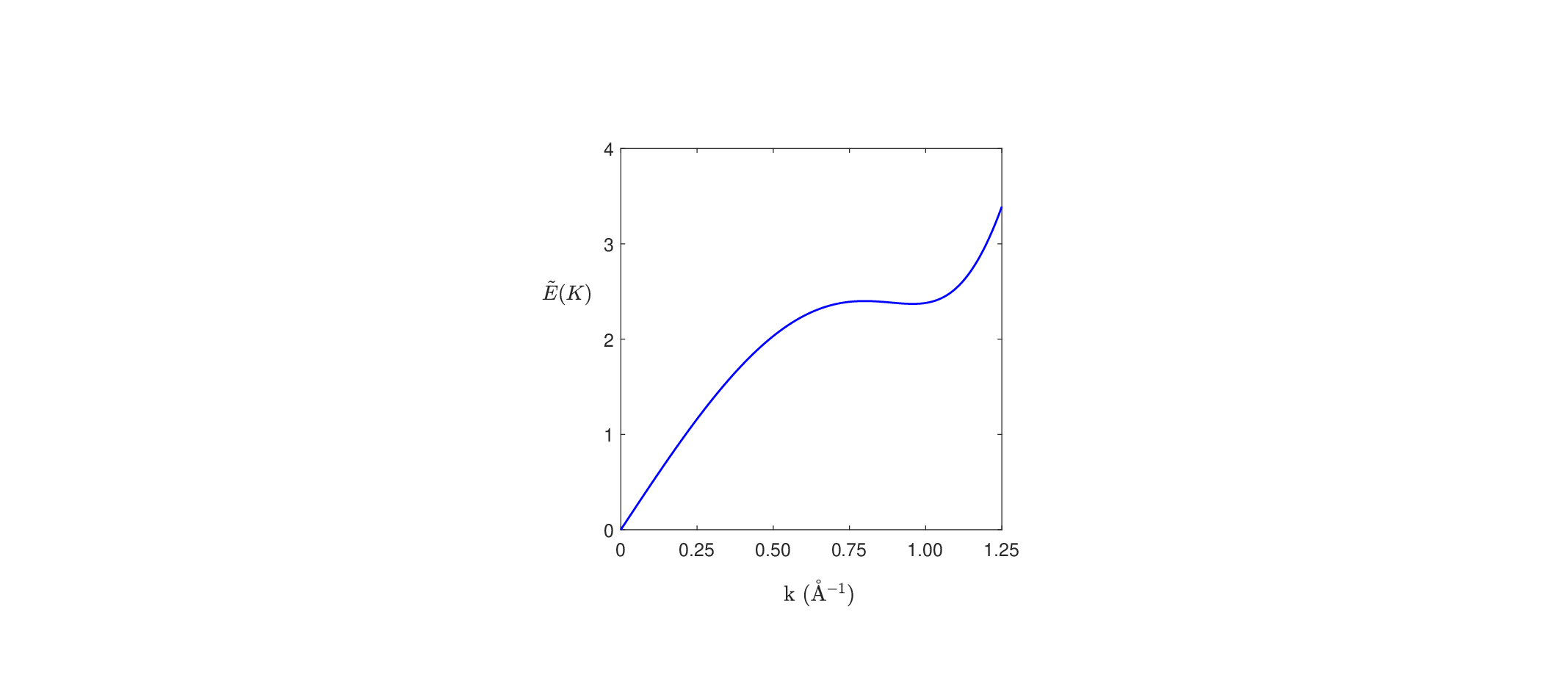}
\caption{Phonon-roton dispersion relation for elementary excitation in superfluid He$^{4}$ film given by Eq. (\ref{81}) for parameters: $c_{s}=63.4~m/s$, $\Delta/k_{B}=2.4$ K and $k_{0}=0.8~\mathring{A}^{-1}$.}
\label{FIG.2.}
\end{figure}
We can also write Eq. (\ref{71}) in the form: 
\begin{equation}
\varepsilon^{2}(p)=\lambda_{1}p^{2}+\lambda_{2}p^{4} +\lambda_{3}p^{6},
\label{73}
\end{equation}
with $\varepsilon(p)=E(k)$ and $p=\hbar k$, then we have 
$\lambda_{1}=b_{1}\hbar^{-2}$, $\lambda_{2}=b_{2}\hbar^{-4}$ and $\lambda_{3}=b_{3}\hbar^{-6}$. We use the following conditions for coefficients $\lambda_{n}$:
\begin{equation}
\left(\frac{d\varepsilon(p)}{dp}\right)_{p=0}=c_{s}, ~~~~ \left(\frac{%
d\varepsilon(p)}{dp}\right)_{p=p_{0}}=0,~~~~ \varepsilon(p_{0})=\Delta,
\label{74}
\end{equation}
with $p_{0}=\hbar k_{0}$. These conditions lead to the following coefficients $\lambda_{n}$: 
\begin{equation}
\lambda_{1}=c_{s}^{2},~~~~ \lambda_{2}=\frac{3\Delta^{2}}{p_{0}^{4}} 
-\frac{2c_{s}^{2}}{p_{0}^{2}},~~~~ \lambda_{3}=\frac{c_{s}^{2}}{p_{0}^{4}}
-\frac{2\Delta^{2}}{p_{0}^{6}},  
\label{75}
\end{equation}
where $\lambda_{3}>0$, $\lambda_{2}<0$ and $\lambda_{1}>0$. 
Using the relations $\lambda_{1}=b_{1}\hbar^{-2}$, $\lambda_{2}=b_{2}\hbar^{-4}$ and $\lambda_{3}=b_{3}\hbar^{-6}$
we have by Eq. (\ref{75}) the coefficients $b_{n}$  as 
\begin{equation}
b_{1}=c_{s}^{2}\hbar^{2},~~~~ b_{2}=\frac{3\Delta^{2}}{k_{0}^{4}}
-\frac{2c_{s}^{2}\hbar^{2}}{k_{0}^{2}},~~~~
b_{3}= \frac{c_{s}^{2}\hbar^{2}}{k_{0}^{4}} -\frac{2\Delta^{2}}{k_{0}^{6}}.  
\label{76}
\end{equation}
Thus, the energy of elementary excitations $E(k)$ has the following explicit form: 
\begin{equation}
E(k)=\left[c_{s}^{2}\hbar^{2}k^{2}+\left(\frac{3\Delta^{2}} {k_{0}^{4}}-%
\frac{2c_{s}^{2}\hbar^{2}}{k_{0}^{2}}\right)k^{4} +\left(\frac{c_{s}^{2}\hbar^{2}}{k_{0}^{4}} -\frac{2\Delta^{2}}{k_{0}^{6}}\right)k^{6}\right]^{1/2}. 
\label{77}
\end{equation}
Moreover, Eqs. (\ref{72}) and (\ref{76}) lead to the following expressions for the coefficients $G$, $\beta$ and $\sigma$ in nonlinear Schr\"{o}dinger type Eqs. (\ref{58}) and (\ref{61}): 
\begin{equation}
G=\frac{mc_{s}^{2}}{\zeta_{0}},~~~~ \beta=\frac{\hbar^{2}}{4m\zeta_{0}}+%
\frac{2mc_{s}^{2}}{\zeta_{0}k_{0}^{2}} -\frac{3m\Delta^{2}}{\zeta_{0}\hbar^{2}k_{0}^{4}},~~~~
\sigma=\frac{mc_{s}^{2}}{\zeta_{0}k_{0}^{4}} -\frac{2m\Delta^{2}}{\zeta_{0}\hbar^{2}k_{0}^{6}}.  
\label{78}
\end{equation}
Equations (\ref{60}) and (\ref{78}) yield the parameters
$\tilde{F_{n}}$ (for $n=0,1,2$) as
\begin{equation}
\tilde{F_{0}}=\frac{mc_{s}^{2}}{n_{0}},~~~~ \tilde{F_{1}}=\frac{\hbar^{2}}{4mn_{0}} +%
\frac{2mc_{s}^{2}}{n_{0}k_{0}^{2}} -\frac{3m\Delta^{2}}{n_{0}\hbar^{2}k_{0}^{4}},~~~~
\tilde{F_{2}}=\frac{mc_{s}^{2}}{n_{0}k_{0}^{4}} -\frac{2m\Delta^{2}}{n_{0}\hbar^{2}k_{0}^{6}}.   
\label{79}
\end{equation}
The third sound velocity in the helium film is defined by equation:
\begin{equation}
c_{s}=\lim_{k\rightarrow 0}=\frac{1}{\hbar}
\frac{\partial E_{\mathbf{k}}}{\partial k}=\sqrt{\frac{n_{0}\,\tilde{F_{0}}}{m}},
\label{80}
\end{equation}
which also yields the relation $\tilde{F_{0}}=mc_{s}^{2}/n_{0}$.
We emphasize that equations for parameters $\tilde{F_{n}}$ in Eq. (\ref{79}) have the same form as coefficients $F_{n}$ in Eq. (\ref{33}). However, the parameters $c_{s}$, $k_{0}$ and $\Delta$ for thin superfluid He$^{4}$ film differ from
appropriate parameters in dense Bose fluid.
We can also write the phonon-roton dispersion equation presented by Eq. (\ref{77}) as
\begin{equation}
\tilde{E}(k)=(Ak^{2}+Bk^{4} +Ck^{6})^{1/2},  
\label{81}
\end{equation}
where $\tilde{E}=E/k_{B}$ and $k=k(\mathring{A}^{-1})$. In this case the coefficients $A$, $B$ and $C$ have the following form: 
\begin{equation}
A=c_{s}^{2}\hbar^{2}k_{B}^{-2}\cdot 10^{20},~~~~ B=3\tilde{\Delta}^{2}\tilde{k}_{0}^{-4} -2c_{s}^{2}\hbar^{2}k_{B}^{-2}\tilde{k}_{0}^{-2}\cdot 10^{20},
~~~~C=c_{s}^{2}\hbar^{2}k_{B}^{-2}\tilde{k}_{0}^{-4}\cdot 10^{20}-2\tilde{\Delta}%
^{2}\tilde{k}_{0}^{-6},
\label{82}
\end{equation}
with $\tilde{\Delta}=\Delta/k_{B}$, and  $k_{0}=k_{0}(\mathring{A}^{-1})$.

We exhibit in Figs. 1 and 2 the phonon-roton dispersion equation given in Eq. (\ref{81}) with the experimental parameters $c_{s}$, $\Delta$ and $k_{0}$ for two different cases Ref. \cite{Rut}. In these figures we use the following experimental parameters for thin superfluid He$^{4}$ film: $c_{s}=59.36~m/s$, $\Delta/k_{B}=5.22$ K, $k_{0}=2~\mathring{A}^{-1}$ (for Fig. 1), and $c_{s}=63.4~m/s$, $\Delta/k_{B}=2.4$ K, $k_{0}=0.8 ~\mathring{A}^{-1}$ (for Fig. 2) respectively.

\section{Conclusion}

In this paper we have derived the extended GP equation for dilute and nearly ideal Bose gas. The derivation is based on the Hartree-Fock time-dependent mean-field theory with an arbitrary intermolecular interaction potential. We have presented in Appendix A the modified Born approximation for dilute Bose gas and the dispersion equation for extended GP equation. 
We have also developed the generalization of this approach leading to novel nonlinear Schr\"{o}dinger type equation for dense liquid helium at low temperatures. The coefficients in this nonlinear Schr\"{o}dinger equation for dense Bose fluid depend on phonon velocity $c$, the wave number $k_{0}$ at roton minimum and the roton minimum energy $\Delta$ of helium Bose fluid. This method yields the unique form for the coefficients in this nonlinear Schr\"{o}dinger equation. The obtained nonlinear Schr\"{o}dinger type equation leads to phonon-roton dispersion relation for elementary excitations in dense Bose fluid. 
We have obtained the quantum term for superfluid component of velocity in two fluid hydrodynamics. This derivation of quantum term for superfluid component of velocity is based on novel nonlinear Schr\"{o}dinger type equation obtained in the Appendix B. This nonlinear Schr\"{o}dinger type equation is derived by the Heisenberg equation for boson annihilation  operator in a Fock space and many-body Hamiltonian for the bosonic system modeling the helium fluid.
We have derived the novel nonlinear Schr\"{o}dinger type equation describing the thin superfluid helium films at low temperatures. This Schr\"{o}dinger type nonlinear equation for thin superfluid He$^{4}$ films depends on two coordinates describing the position on helium surface.
It is show that the coefficients in this Schr\"{o}dinger type nonlinear equation for thin superfluid He$^{4}$ films are uniquely defined by the third sound velocity $c_{s}$ in helium film, the wave number $k_{0}$ at roton minimum $\Delta$ and the roton minimum energy $\Delta$. This nonlinear Schr\"{o}dinger type equation  leads to phonon-roton dispersion relation for elementary excitations in thin superfluid He$^{4}$ films.  
We anticipate that obtained in this paper the nonlinear Schr\"{o}dinger type equations and dispersion relations for the elementary phonon-roton excitation in dense Bose fluid and superfluid He$^{4}$ films can find various applications.

\appendix

\section{Modified Born approximation for extended GP equation}

In this Appendix A we present the results obtained
in the papers \cite{A14,A15} leading to modified Born approximation for the extended GP Eq. (\ref{16}) describing dilute Bose gas. We note that this equation
is applicable under two conditions: $\sqrt{a_{0}^{3}n}\ll 1$ and $ka_{0}\ll 1$ where $k$ is the wave number, and the second condition means that the interaction of particles is described by the s-scattering waves. We note that the GP equation follows from Eq. (\ref{16}) if we take $D_{1}=0$ and $D_{2}=0$ in 
Eq. (\ref{17}).

We use in our approach the standard definition for the s-wave scattering length: 
\begin{equation}
\frac{1}{a_{0}}=-\lim_{k\rightarrow 0}
(k\cot\,\delta_{0}(k)).	
\label{a1}
\end{equation}
where $\delta_{0}(k)$ is the phase shift of the s-scattering wave function and $k$ is the wave number. This equation can also be written as
\begin{equation}
a_{0}=\lim_{k\rightarrow 0}\frac{m}{\hbar^{2}}
\int_{0}^{\infty}U(r)\phi_{k}(r)rdr,
\label{a2}
\end{equation}
where $U(r)$ is the scattering potential and $\phi_{k}(r)$ is an exact wave function defined in the scattering theory with appropriate boundary conditions. When the Born approximation is valid the wave function $\phi_{k}(r)$ in equation (\ref{a2}) can be replaced by the wave function of a free particle in the form $\phi_{k}^{(0)}(r)=k^{-1}\sin kr$. However, for many scattering potentials $U(r)$ this approximation is meaningless because the integral in equation (\ref{a2}) diverges.

In the modified Born approximation (MBA) \cite{A14,A15} the wave function in equation (\ref{a2}) is instead approximated by
\begin{equation}
\phi_{k}(r)=\theta(r-a_{0})k^{-1}\sin kr, 
\label{a3}
\end{equation}
where $\theta(r)$ is the Heaviside unit step function. Thus, the wave function in equation (\ref{a3}) is zero for $r<a_{0}$; the region $r<a_{0}$
is unattainable for slow particles ($k\rightarrow 0$) because
the cross-section for s-scattering waves is
$\sigma_{s}=4\pi a_{0}^{2}$. Equations (\ref{a2}) and (\ref{a3}) lead to an equation for the s-wave scattering length as
\begin{equation}
a_{0}=\frac{m}{\hbar^{2}}\int_{a_{0}}^{\infty}U(r)r^{2}dr=
\frac{m}{\hbar^{2}}\int_{0}^{\infty}\tilde{U}(r)r^{2}dr,
\label{a4}
\end{equation}
where the effective potential $\tilde{U}(r)$ incorporates the cutoff: $\tilde{U}(r)=0$ for $r<a_{0}$ and $\tilde{U}(r)=U(r)$ for $r\geq a_{0}$.
We have defined in Eq. (\ref{10}) an effective interatomic potential $U_{a}(r)$ with cutoff as: $U_{a}(r)=0$ for $r<a$ and $U_{a}(r)=U(r)$ otherwise. Thus, for dilute gas the cutoff parameter is $a_{0}=a$, and we have the relation $U_{a}(r)=\tilde{U}(r)$. This means that for dilute gas the coupling parameters $G_{0}$, $D_{1}$ and $D_{2}$ in extended GP Eq. (\ref{16}) have the cutoff parameter $a=a_{0}$ 

We have found that for dilute gas the cutoff parameter is given by s-wave scattering length $a_{0}$, which yields the coupling parameter $G_{0}$ in Eq. (\ref{10}) as
\begin{equation}
G_{0}=4\pi\int_{a_{0}}^{\infty}U(r)r^{2}dr,
\label{a5}
\end{equation}
with $a=a_{0}$. Equations (\ref{a4}) and (\ref{a5}) yield the relation $a_{0}\hbar^{2}/m=G_{0}/4\pi$ leading to the following equation for coupling parameter $G_{0}$:
\begin{equation}
G_{0}=\frac{4\pi a_{0}\hbar^{2}}{m}.  
\label{a6}
\end{equation}

In the particular case when the potential $U(r)$ is given by the Lennard-Jones potential the coupling parameter $G_{0}$ follows from Eq. (\ref{a5}) as
\begin{equation}
G_{0}=\frac{16\pi}{9}\epsilon r_{0}^{3}\left[\left(\frac{r_{0}}{a_{0}}\right)^{9}
-3\left(\frac{r_{0}}{a_{0}}\right)^{3}\right].
\label{a7}
\end{equation}

Equations (\ref{a6}) and (\ref{a7}) lead to the following equation for the s-wave scattering length: 
\begin{equation}
a_{0}=\frac{4m\epsilon r_{0}^{3}}{9\hbar^{2}}\left[\left(\frac{r_{0}}{a_{0}}\right)^{9}-3\left(\frac{r_{0}}{a_{0}}\right)^{3}\right].
\label{a8}
\end{equation}
Thus, Eq. (\ref{a8}) yields a fifth-order algebraic equation,
\begin{equation}
X^{5}-3X^{2}-X_{0}=0, 
\label{a9}
\end{equation}
where $X$ and $X_{0}$ are given as
\begin{equation}
X=\left(\frac{r_{0}}{a_{0}}\right)^{2},~~~~
X_{0}=\frac{9\hbar^{2}}{4m\epsilon r_{0}^{2}}.
\label{a10}
\end{equation}
The solution of Eq. (\ref{a9}) with the parameters given for He$^{4}$ [$\epsilon/k_{B}=10.6\,K$ and $r_{m}=2.98\, \mathring{A}$] leads to s-wave scattering length $a_{0}=2.20\, \mathring{A}$. 

In Sec. II, we have derived the novel nonlinear Schr\"{o}dinger type equation (\ref{16}) generalizing the GP equation for dilute Bose gas. In this derivation of Eq. (\ref{16})
we have neglected the remainder $R_{5}(\mathbf{x^{\prime}},t)$ in the Taylor series given by  Eq. (\ref{7}). 
A reasonable criterion for such approximation follows from
Eq. (\ref{17}) for the nonlinear interaction potential: 
\begin{equation}
|D_{2}\nabla^{4}n(\mathbf{x},t)|\ll 
|G_{0}n(\mathbf{x},t)+D_{1}\nabla^{2}n(\mathbf{x},t)|.
 \label{a11}
\end{equation}
This criterion can be used if the variation of density $n(\mathbf{x},t)$ and the functions $\nabla^{2}n(\mathbf{x},t)$ and $\nabla^{4}n(\mathbf{x},t)$ are known.
Hence, first one should find numerically (or analytically) the solution of nonlinear Schr\"{o}dinger equation (\ref{16}) for appropriate physical problem. In this case all terms in Eq. (\ref{a11}) can be calculated numerically or analytically. Nevertheless the above criterion is not a necessary condition for the validity of extended GP equation. 

In the conclusion of this Appendix A we present the dispersion equation for the extended GP Eq. (\ref{16}): 
\begin{equation}
\hbar\omega_{\mathbf{k}}=\left[\left(\frac{\hbar^{2}k^{2}}{2m}\right)^{2}+n_{0}\left(\frac{\hbar^{2} k^{2}}{m}\right)
(G_{0}-D_{1}k^{2}+D_{2}k^{4})\right]^{1/2}.
\label{a12}
\end{equation}
The derivation of dispersion equation (\ref{a12}) for the extended GP Eq. (\ref{16}) is based on the method given in
Sec. III. This dispersion equation also follows by the change $F_{0}\mapsto G_{0}$, $F_{1}\mapsto D_{1}$ and $F_{2}\mapsto D_{2}$ in Eq. (\ref{26}). Equation (\ref{a12}) shows that the parameters $G_{0}$, $D_{1}$ and $D_{2}$  define the terms proportional to $k^{2}$, $k^{4}$ and $k^{6}$ for the function $\omega_{\mathbf{k}}^{2}$.

\section{Schr\"{o}dinger type equation for superfluid helium in two fluid hydrodynamics}

In this Appendix B we consider the derivation of nonlinear Schr\"{o}dinger type equation describing the superfluid helium component in two fluid hydrodynamics. This nonlinear Schr\"{o}dinger type equation leads to quantum correction in two fluid hydrodynamics developed by  Landau \cite{Lan,Pat}.
Our approach is based on the Heisenberg equations for boson annihilation and creation field operators $\hat{\Psi}(\mathbf{x},t)$ and $\hat{\Psi}^{\dagger}(\mathbf{x},t)$ in a Fock space. We define the wave function $\Psi(\mathbf{x},t)$ and 
the operator $\hat{\eta}(\mathbf{x},t)$ in the form,
\begin{equation}
\Psi(\mathbf{x},t)=\langle\hat{\Psi}(\mathbf{x},t)\rangle,
~~~~\hat{\eta}(\mathbf{x},t)=\hat{\Psi}(\mathbf{x},t)
-\langle\hat{\Psi}(\mathbf{x},t)\rangle.	
\label{b1}
\end{equation}
This equation yields the relations: $\hat{\Psi}(\mathbf{x},t)=\Psi(\mathbf{x},t)+\hat{\eta}
(\mathbf{x},t)$ and $\langle\hat{\eta}(\mathbf{x},t)\rangle$=0,
where $\langle...\rangle=$Tr$(...\rho_{0})$ is the averaging with the density operator $\rho_{0}$ at an initial time $t=0$. 
The above definitions and relations lead to equation:
\begin{equation}
\langle\hat{\Psi}^{\dagger}
(\mathbf{x},t)\hat{\Psi}(\mathbf{x},t)\rangle
=\nu_{s}(\mathbf{x},t)+\nu_{n}(\mathbf{x},t),
\label{b2}
\end{equation}
where the superfluid and normal densities of helium fluid (below lambda transition) are given
as $\nu_{s}(\mathbf{x},t)=|\Psi(\mathbf{x},t)|^{2}$ and $\nu_{n}(\mathbf{x},t)=\langle\hat{\eta}^{\dagger}
(\mathbf{x},t)\hat{\eta}(\mathbf{x},t)\rangle$.

The modified Hamiltonian $\hat{H}-\mu\hat{N}$ for the boson system with the interatomic potential can be written in the form:
\begin{equation}
\hat{H}-\mu\hat{N}=\int\hat{\Psi}^{\dagger}(\mathbf{x})
\left(-\frac{\hbar^{2}}{2m}\nabla^{2}-\mu\right)
\hat{\Psi}(\mathbf{x})d^{3}x+\frac{1}{2}\int\int
\hat{\Psi}^{\dagger}(\mathbf{x})\hat{\Psi}^{\dagger}
(\mathbf{x^{\prime}})U(|\mathbf{x}-\mathbf{x^{\prime}}|)
\hat{\Psi}(\mathbf{x^{\prime}})\hat{\Psi}(\mathbf{x})
d^{3}xd^{3}x^{\prime}.
\label{b3}
\end{equation}
This Hamiltonian yields the Heisenberg equation for the
time-dependent field operator $\hat{\Psi}(\mathbf{x},t)$ as
\begin{equation}
i\hbar\frac{\partial}{\partial t}\hat{\Psi}(\mathbf{x},t)
=-\frac{\hbar^{2}}{2m}\nabla^{2}\hat{\Psi}(\mathbf{x},t)
-\mu\hat{\Psi}(\mathbf{x},t)
+\hat{U}(\mathbf{x},t)\hat{\Psi}(\mathbf{x},t), 
\label{b4}
\end{equation}
where the interaction operator $\hat{U}(\mathbf{x},t)$ has the form:
\begin{equation}
\hat{U}(\mathbf{x},t)=\int U(|\mathbf{x}-\mathbf{x^{\prime}}|)\hat{\Psi}^{\dagger}
(\mathbf{x^{\prime}},t)\hat{\Psi}(\mathbf{x^{\prime}},t)
d^{3}x^{\prime}.
\label{b5}
\end{equation}
The averaging of Eq. (\ref{b4}) with the density operator $\rho_{0}$ leads to the following nonlinear Schr\"{o}dinger type equation:
\begin{equation}
i\hbar\frac{\partial}{\partial t}\Psi(\mathbf{x},t)
=-\frac{\hbar^{2}}{2m}\nabla^{2}\Psi(\mathbf{x},t)
-\mu\Psi(\mathbf{x},t)
+\langle\hat{U}(\mathbf{x},t)\hat{\Psi}
(\mathbf{x},t)\rangle, 
\label{b6}
\end{equation}
where the term $\langle\hat{U}(\mathbf{x},t)\hat{\Psi}
(\mathbf{x},t)\rangle$ is given by
\begin{equation}
\langle\hat{U}(\mathbf{x},t)\hat{\Psi}
(\mathbf{x},t)\rangle=\langle\hat{U}(\mathbf{x},t)
[\Psi(\mathbf{x},t)+\hat{\eta}(\mathbf{x},t)]\rangle
=\langle\hat{U}(\mathbf{x},t)\rangle\Psi(\mathbf{x},t)
+\langle\hat{U}(\mathbf{x},t)\hat{\eta}
(\mathbf{x},t)\rangle.
\label{b7}
\end{equation}
We note that the term 
$\langle\hat{U}(\mathbf{x},t)\hat{\eta}
(\mathbf{x},t)\rangle$ can be neglected approximately in Eq. (\ref{b7}). In this case the nonlinear Eq. (\ref{b6}) can be written in the form: 
\begin{equation}
i\hbar\frac{\partial}{\partial t}\Psi(\mathbf{x},t)
=\left\{-\frac{\hbar^{2}}{2m}\nabla^{2}-\mu
+\langle\hat{U}(\mathbf{x},t)\rangle\right\}
\Psi(\mathbf{x},t), 
\label{b8}
\end{equation}
where the nonlinear potential $\langle\hat{U}(\mathbf{x},t)\rangle$ [with $\nu_{s}(\mathbf{x},t)=|\Psi(\mathbf{x},t)|^{2}$ and $\nu_{n}(\mathbf{x},t)=\langle\hat{\eta}^{\dagger}
(\mathbf{x},t)\hat{\eta}(\mathbf{x},t)\rangle$]
is given as
\begin{equation}
\langle\hat{U}(\mathbf{x},t)\rangle=\int U(|\mathbf{x}-\mathbf{x^{\prime}}|)(\nu_{s}
(\mathbf{x}^{\prime},t)+\nu_{n}(\mathbf{x}^{\prime},t))
d^{3}x^{\prime}.
\label{b9}
\end{equation}

\section{Madelung equations}

Madelung equations are the alternative formulation of the
one particle linear Schr\"{o}dinger equation in terms of hydrodynamical variables \cite{M}.  This paper was one of numerous early attempts at different approaches to quantum mechanics. The Madelung equations are similar to the Bohm formulation, which represents the Schrödinger equation as a quantum Hamilton–Jacobi equation \cite{Bom}. In both cases the hydrodynamic interpretations are not equivalent to accepted interpretation of quantum mechanics.
The derivation of Madelung equations starts with one particle linear Schr\"{o}dinger equation 
where the wave function has the form
$\psi(\mathbf{x},t)=\sqrt{\rho(\mathbf{x},t)}
\exp\left(\frac{i}{\hbar}  S(\mathbf{x},t)\right)$ and
the probability density and velocity are given by relations: $\rho(\mathbf{x},t)=|\psi(\mathbf{x},t)|^{2}$ and $\mathbf{v}(\mathbf{x},t)=\frac{1}{m}\nabla S(\mathbf{x},t)$. 
The interpretation of $\mathbf{v}(\mathbf{x},t)$ as "velocity`` should not be taken too literally because a simultaneous precision measurement of position and velocity would necessarily violate the uncertainty principle which is a fundamental concept in quantum mechanics \cite{Sak}.

The linear Schr\"{o}dinger equation for one particle with the potential $V(\mathbf{x},t)$ can be written in terms of
density $\rho$ and "velocity`` $\mathbf{v}$ as 
a system of two coupled partial differential equations Ref. \cite{M}:
\begin{equation}
\partial_{t}\rho+\nabla\cdot (\rho\mathbf{v})=0,~~~~
\partial_{t}\mathbf{v}+(\mathbf{v}\cdot\nabla)\mathbf{v}
=-\frac{1}{m}\nabla \left(V(\mathbf{x},t)+Q(\mathbf{x},t)\right),
\label{c1}
\end{equation}
where the function $Q(\mathbf{x},t)$ 
("quantum potential``) is given by
\begin{equation}
Q(\mathbf{x},t)=-\frac{\hbar^{2}}{2m}
\frac{\nabla^{2}\sqrt{\rho(\mathbf{x},t)}}
{\sqrt{\rho(\mathbf{x},t)}}.	
\label{c2}
\end{equation}
The Madelung equations derived for alternative formulation of  one particle linear Schr\"{o}dinger equation are similar to the Euler equations of fluid dynamics. These equations significantly differ from Eqs. (\ref{40}) and (\ref{45}) which describe the dynamics of superfluid component of density and velocity in the Landau two fluid hydrodynamics with quantum term $\nabla R$. 
We note that equations for the superfluid density $\rho_{s}$ and velocity $\mathbf{v}_{s}$ are found by nonlinear Schr\"{o}dinger type equation (\ref{37}) for the wave function $\Psi(\mathbf{x},t)=\langle\hat{\Psi}(\mathbf{x},t)\rangle$ where $\hat{\Psi}(\mathbf{x},t)$ is the boson annihilation operator.

\end{document}